\documentclass{aa}
\usepackage[varg]{txfonts}
\usepackage{txfonts}
\usepackage{color}
\usepackage{graphicx}

\newcommand{\avg}[1]{\ensuremath{\langle #1 \rangle}}

\begin{document}
\title{Small-scale dynamo in cool stars}
\subtitle{III. Changes in the photospheres of F3V to M0V stars}
\author{Tanayveer Bhatia\inst{1}\fnmsep\thanks{\email{bhatia@mps.mpg.de}}\and
    Robert Cameron\inst{1}\and
    Hardi Peter\inst{1}\and
    Sami Solanki\inst{1}}
    \institute{Max-Planck-Institut f\"ur Sonnensystemforschung, Justus-von-Liebig Weg 3, 37077 Göttingen}
\date{Received / Accepted }
\abstract
    {Some of the quiet solar magnetic flux could be attributed to a small-scale dynamo (SSD) operating in the convection zone. An SSD operating in cool main-sequence stars is expected to affect the atmospheric structure, in particular the convection, and should have observational signatures.}
    {We aim to investigate the distribution of SSD magnetic fields as well as their effect on bolometric intensity characteristics, vertical velocity and spatial distribution of kinetic energy (KE) and magnetic energy (ME) in the lower photosphere of different spectral types.}
    {We analyse the SSD and purely hydrodynamic simulations of the near surface layers of F3V, G2V, K0V and M0V stars. We compare the time-averaged distributions and power spectra in SSD setups relative to the hydrodynamic setup. Properties of the individual magnetic fields are also considered.}
    {PDFs of field strength at the $\tau=1$ surface are quite similar for all cases. The M0V star displays the strongest fields, but relative to the gas pressure, the fields on the F3V star reach the largest values. All stars display an excess of horizontal field relative to vertical field in the middle photosphere, with this excess becoming increasingly prominent towards later spectral types. These fields result in a decrease in upflow velocities, slightly smaller granules as well as the formation of bright points in intergranular lanes. The spatial distribution of KE and ME is also similar for all cases, implying a simple pressure scale height proportionality of important scales.}
    {SSD fields have rather similar effects on the photospheres of cool main-sequence stars, namely, a significant reduction in convective velocities as well as a slight reduction in granule size, and concentration of field to kG levels in intergranular lanes associated with the formation of bright points. The distribution of field strengths and energies is also rather similar.}
    
\maketitle 

\section{Introduction}

Magnetism in cool stars is ubiquitous. In addition, a significant number of cool stars show a solar-like activity cycle \citep{Wilson1978}. The magnetic fields associated with these cycles are expected to arise from a large-scale dynamo operating in the interiors of cool stars \citep{BranSub2005,Char2014}. However, there is also an additional, cycle-independent component of stellar fields, the quiet-star magnetism. From detailed observations of the quiet Sun, (see, e.g., reviews by \citet{QS1993,ssfields_dewijn2009,QS2011,QSreview2019}) as well as state-of-the-art simulations \citep{VogSch2007,Rempel2014}, this component was realised to be substantial and could, in part, be explained by invoking a small-scale dynamo (SSD) mechanism which would amplify magnetic fields via turbulent motions of the plasma. In fact, recent global SSD simulations \citep{Hotta2021,Hotta2022} showed that the field generated can be significantly super-equipartition at small scales in the deep convection zone, being strong enough to affect the meridional circulation and the differential rotation profile. For stars other than the sun, the influence of SSD fields on quiet star phenomena like granulation, pressure oscillations and basal chromospheric activity remain yet to be studied.

Quiet star magnetic fields also have the potential to affect interpretation of radial velocity (RV) observations of stellar spectra. RV measurements allow detection of exoplanets by accounting for Doppler shifts in stellar spectral lines due to a gravitationally-induced "wobble" caused by a planet's orbital motion. However RVs can be affected by stellar magnetism. Starspots can affect RVs in multiple ways, e.g. via the Evershed effect \citep{sami2003,RimMar2006}, while faculae reduce the granular blueshift \citep{BraSol1990}. Granulation, whose presence is not just visible in short time-scale "noise", but also as a net blueshift in photospheric spectral lines of most solar-like stars \citep{Dravins1987}, could also potentially affect RVs. For a comprehensive list of factors potentially affecting high precision RV measurements, see Table A-4 in \citet{eprv}. \citet{ShpBro2011} demonstrated the impact convective blueshift can have on RV measurements during transits at m/s accuracy level via a simple model. With RV measurements reaching sub-m/s precision, allowing detection of Earth-like rocky exoplanets, it becomes imperative to understand the sources of stellar "noise" properly, including the contribution from magnetic fields. 

Stellar light curves also show variation on the order of granulation timescales. The amplitude of variations over such short timescales are well-correlated with the stellar surface gravity \citep{Bastien2013,Bastien2016}, allowing an independent estimation of the latter (as opposed to asteroseismic measurements). In addition, \citet{Sasha2017} showed that, for the Sun, the total solar irradiance (TSI) could be reliably reconstructed just from the consideration of granulation noise from simulations and solar magnetograms in a forward model. This is encouraging for modelling stellar variability at shorter timescales.

In our recent work (\citet{paper1}, hereafter Paper I), we showed that SSD magnetic fields significantly reduce the convective velocities and can be strong enough in earlier spectral types (F-star and earlier) to even affect the stratification and scale heights near the surface (which could influence scales of granulation). In addition, it was also shown that changing the metallicity also leads to a difference in SSD-associated field strengths and properties of momentum transport near the surface \citep{vero2022}. Hence, it becomes imperative to understand the effect of SSD fields on the photospheres of different stars.

In this paper, we describe the distribution of photospheric quiet-star magnetic fields as expected to arise from an SSD mechanism operating in the near-surface convection of late-type dwarfs. We also look at the effects of this magnetic field on the velocities, the bolometric intensity as well as the energy distribution in the photospheres of these stars.

\section{Methods}\label{sec:methods}

We use the models described in Paper I, namely, the purely hydrodynamic (HD) models and the models with SSD fields. The setup, number of snapshots, time range etc. are similar to those in Paper 1. Briefly, we consider four sets of a local box-in-a-star simulation of F3V, G2V, K0V and M0V stars, with each set consisting of a time series of an HD run and an SSD run. The boxes have 512 grid points in both the horizontal directions and 500 grid points in the vertical direction. The resolution (and the physical size) is such that all simulations cover a similar number of pressure scale heights ($n_{H_p} = \log (p_{\mathrm{gas}}/p_{\mathrm{gas}(\avg{\tau}=1)})$) and horizontally scaled to maintain the aspect ratio. The G2V star is used as a reference, with $n_{H_p} \sim 7.5$ below the surface and $\sim 6 - 8$ above the surface (corresponding to 4 Mm below, 1 Mm above). The horizontal extent is 9 Mm $\times$ 9 Mm \footnote{The K-star SSD and HD setups were rerun for this paper with a taller atmosphere so as to have a similar $n_{H_p}$ above the surface for all cases. The new time series was used to redo all analyses mentioned in Paper I and the resulting plots were practically the same.}. This corresponds to a resolution of approximately 17.6(10) km in the horizontal (vertical) direction. For the F, K and M-stars, the corresponding resolutions are approximately 45(26), 8.2(4.6) and 4(2.3) km, respectively. The stellar type is determined by specifying the entropy of inflows (effectively setting the $T_{\rm eff}$) and by setting a constant gravitational acceleration in the $z-$direction. We note that the upper boundary condition for magnetic field is set to be vertical. We simply made a choice between vertical and potential upper boundary\footnote{A test run with potential upper boundary for the G2V star resulted in more horizontal fields above the surface, and a small increase in the magnitude of magnetic field (possibly a consequence of more low-lying loops available for recirculation and amplification). At and below the $\tau=1$ surface, there was practically no change in the thermodynamic structure.}.  We refer the reader to Paper 1 for further details about the setup.

Before we start describing the results, we provide some information on the meaning of symbols and conventions we follow. All averages over time are denoted by an overline $\overline{q}$. All averages over space of 2D data are denoted by angular brackets $\avg{q}$. The standard deviation for bolometric intensity in a single snapshot is calculated as $\sigma_I = \sqrt{\avg{(I-\avg{I})^2}}$. The calculation of spatial power spectra is covered in Appendix \ref{app:ibol_ps}. All plots have error bars corresponding to standard error $\epsilon = \sigma/\sqrt{N}$, where $N$ is the number of snapshots and $\sigma$ is the standard deviation over time averages. The colour coding is the same as that in Paper I, with blue for the F3V, black for the G2V, green for the K0V, and red for the M0V star. Dashed lines with lighter corresponding colours refer to the hydrodynamic case, unless stated otherwise.

\section{Results}
\subsection{Distribution of the magnetic fields}\label{sec:results:bpdf}

\begin{figure}[h]
    \resizebox{\hsize}{!}
    {\includegraphics{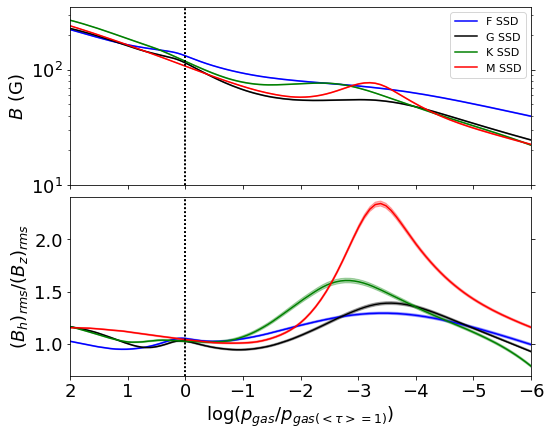}}
    \caption{Magnetic field characteristics. \textit{Top}: Horizontally averaged magnitude of magnetic field. \textit{Bottom}: Horizontally averaged inclination of magnetic field, defined as $\sqrt{\avg{B_x^2+B_y^2}/\avg{B_z^2}}$. The horizontal axis is the number of pressure scale heights below the surface (positive is below, negative is above).}
    \label{fig:b_charac}
\end{figure}

\begin{figure}[h]
    \resizebox{\hsize}{!}
    {\includegraphics{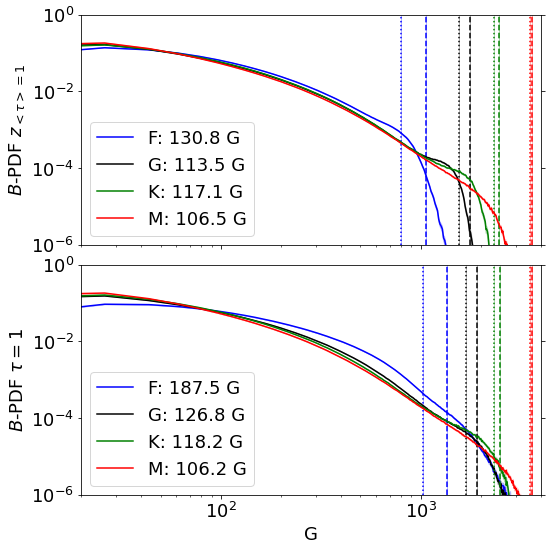}}
    \caption{PDF of the strength of magnetic field $B$ for the SSD F- (\textit{blue}), G- (\textit{black}), K- (\textit{green}) and M-star (\textit{red}). \textit{Top}: PDF of $B$ calculated for the geometric surface $z_{\avg{\tau}=1}$ corresponding to the height at which $\avg{\tau}=1$. \textit{Bottom}: Same plot as the top, but for the iso-$\tau=1$ surface. The vertical lines correspond to the pressure equipartition field $B_{\rm eqp,gas} = \sqrt{8\pi (p_{\mathrm{gas}})}$ (dotted) and $B_{\rm eqp,tot} = \sqrt{8\pi (p_{\mathrm{gas}}+\rho v_h^2})$ (dashed), for each case (see Sect. \ref{sec:discuss:bpdf} for details on the description of $B_{\rm eqp,tot}$). The labels contain the mean value for each case in gauss.}
    \label{fig:bpdf}
\end{figure}

\begin{table*}[h]
    \caption{\textit{From left to right:} Characteristic of magnetic field near the surface for each stellar type: average field strength ($\overline{B}$), average vertical field strength ($\overline{|B_z|}$), average horizontal field strength ($\overline{|B_h|}$) and the percent fraction of area occupied by kilogauss fields. ($\overline{A_{B>1\mathrm{kG}}/A_{\mathrm{tot}}}$). All values are averages over time along with 1$\sigma$ standard deviation.}
    \label{tab:b_charac}
    \centering
    \begin{tabular}{c|c c | c c |c c |c c}
        \hline\hline
        Simulation & \multicolumn{2}{c|}{$\overline{B}$ [G]} & \multicolumn{2}{c|}{$\overline{|B_z|}$ [G]} &  \multicolumn{2}{c|}{$\overline{B_h}$ [G]} &
        \multicolumn{2}{c}{$\overline{A_{B>1\mathrm{kG}}/A_{\mathrm{tot}}}$ [\%]}\\
         & $z_{\avg{\tau}=1}$ & $\tau=1$ & $z_{\avg{\tau}=1}$ & $\tau=1$ & $z_{\avg{\tau}=1}$ & $\tau=1$ & $z_{\avg{\tau}=1}$ & $\tau=1$ \\
        \hline
        F3V & 130 $\pm$ 10 & 187 $\pm$ 14 &	66 $\pm$ 6 & 94 $\pm$ 8 &	98 $\pm$ 7 & 143 $\pm$ 11 &	0.07 $\pm$ 0.06 & 0.80 $\pm$ 0.31\\
        G2V & 113 $\pm$ 10 & 126 $\pm$ 11 &	59 $\pm$ 5 & 66 $\pm$ 6 &	85 $\pm$ 8 & 95 $\pm$ 9 &	0.46 $\pm$ 0.14 & 0.47 $\pm$ 0.14\\
        K0V & 117 $\pm$ 9 & 118 $\pm$ 9 &	60 $\pm$ 4 & 60 $\pm$ 4 &	89 $\pm$ 7 & 90 $\pm$ 7 &	0.53 $\pm$ 0.12 & 0.50 $\pm$ 0.11\\
        M0V & 106 $\pm$ 6 & 106 $\pm$ 6 &	55 $\pm$ 3 & 55 $\pm$ 3 &	81 $\pm$ 5 & 81 $\pm$ 4 &	0.43 $\pm$ 0.17 & 0.42 $\pm$ 0.16\\
        \hline
    \end{tabular}
\end{table*}

Fig. \ref{fig:b_charac} shows the horizontally-averaged magnitude (top panel) and horizontally-averaged inclination (bottom panel) of the magnetic field near the surface. All cases show similar value of magnetic field strength, except for the F-star, which shows a somewhat higher value at and above the surface (which is marked by the dotted vertical line). The field inclination shows the field becoming more horizontal for all cases as one goes higher up in the atmosphere. Between $-2>\log(p_{\mathrm{gas}}/p_{\mathrm{gas}(\avg{\tau}=1)})>-4$, all cases show a peak in $B_h/B_z$, which probably corresponds to low lying magnetic field loops. Higher up, $B_h/B_z$ tends towards zero, in accordance with the upper boundary condition of a vertical field.

Table \ref{tab:b_charac} lists the magnetic field characteristics for all cases for $\avg{\tau}=1$ horizontal slice as well as for $\tau=1$ iso-surface\footnote{The $\tau=1$ iso-surface refers to the surface where $\tau=1$ for each vertical column in the 3D cube. The data points are calculated by interpolating logarithmically against the corresponding $\tau$ column to where $\tau=1$.}. Here, $\tau$ refers to the optical depth corresponding to a reference wavelength of 500 nm. The $\tau=1$ iso-surface, hence, provides an observational point of view for understanding the results. We also consider the horizontal slice because the thermodynamic stratification is expected to be quite uniform in the horizontal direction, allowing a better understanding of the physics of the magnetic field distribution. The columns show that for the horizontal slice, the average field strength is quite similar (100 to 130 G) for all cases, but the value increases significantly for the F-star (almost 190 G) and G-star (almost 130 G) if the $\tau=1$ iso-surface is considered. For the other cases, the change is almost within the standard deviation and decreases with $T_{\rm eff}$. In addition, the area fraction of kG fields for the F-star is around 0.8\% for the $\tau=1$ surface, whereas for the $\avg{\tau}=1$ horizontal slice, it is almost zero. We note here that this result is a consequence of the field strength corresponding to pressure equipartition (including the $\rho v_h^2$ term) at this height being just about one kilogauss (kG); this result is discussed in detail in Sect. \ref{sec:discuss:bpdf}. For the other cases, the area fraction is roughly 0.5\% both ways.

Fig. \ref{fig:bpdf} shows the probability density function (PDF) of the magnitude of the magnetic field for the horizontal slice at $\avg{\tau}=1$ (top panel) and for the $\tau=1$ iso-surface (bottom panel). All stars show a rather similar distribution of photospheric magnetic fields $B$, with most of the field between 20 to 100 G (slightly higher for the F-star) and a more rapid drop-off ("knee") in the kG regime. For the horizontal slice, the field strength for the kG-knee shows an inverse trend with $T_{\mathrm{eff}}$. These kG fields form mostly in the downflow lanes (see appendix \ref{app:add_plots} for separate PDFs of magnetic field in upflows and downflows). The vertical dotted and dashed lines mark the field strength corresponding to equipartition with $p_{\rm gas}$ (that is, $B_{\rm eqp,gas}$) and $p_{\rm gas}+\rho v_h^2$ (that is, $B_{\rm eqp,tot}$), respectively, for all the cases. We note that there seems to be a trend in the location of the kG-knee and the equipartition field strength. We also note that, if compared against just gas pressure, the F-star seems to have super-equipartition fields. This tendency towards having super-equipartition fields relative to $p_{\rm gas}$ decreases towards later spectral types, with the M-star fields being decidedly sub-equipartition. We discuss this relation further in Sect. \ref{sec:discuss:bpdf}. Briefly, we expect the strongest fields to be situated in intergranular lanes, with the strength of the field being such that it balances against the external pressure.

When one considers the $\tau = 1$ iso-surface (bottom panel), the distribution of fields is roughly similar and does not show the trend for kG fields from the $\avg{\tau}=1$ slice. In addition, the field strengths for the F-star are generally higher. For the fields concentrated by flux expulsion, i.e. the fields in equipartition with the flows (around a few 100 G), this is due to a depression in the $\tau=1$ iso-surface in the intergranular lanes. As opacity in photospheres of cool stars depends steeply on temperature, the $\tau=1$ surface generally dips below $z_{\avg{\tau}=1}$ in the cooler intergranular lanes. For the kG fields, however, the dominant mechanism is evacuation of plasma in the flux tubes leading $\tau=1$ forming deeper below and radiation escaping from lower levels. This depression in optical depth is somewhat similar to the Wilson depression (WD) \citep{wilson1774} of the optical surface in sunspots. Because of increasing pressure due to stratification and conservation of flux, the field strength is stronger deeper down. The magnitude of this depression scales with $T_{\mathrm{eff}}$ as well as pressure scale height \citep{beeck3}, and is the strongest for the F-star and weakest for the M-star. These results are consistent with simulations of weak stellar magnetism by \citet{salhab2018}, where the authors also reported similar values of kG field concentrations for all cases as well as the scaling of Wilson depression with pressure scale height. The factors influencing the kG field distribution are discussed later in the context of convective collapse \citep{spruit1979} as well as pressure balance in individual flux concentrations in Sect. \ref{sec:discuss:bpdf}.

\subsection{Bolometric intensity}\label{sec:results:ibol}

In Paper 1, we considered the changes in thermodynamic structure due to SSD-generated magnetic fields. We showed that these changes resulted in a decrease in density scale height $H_{\rho}$, as well as in convective velocities, near the stellar surface. Here, we consider how these changes affect the intensity structure.

\begin{figure*}[h]
    \centering
    \includegraphics[width=17cm]{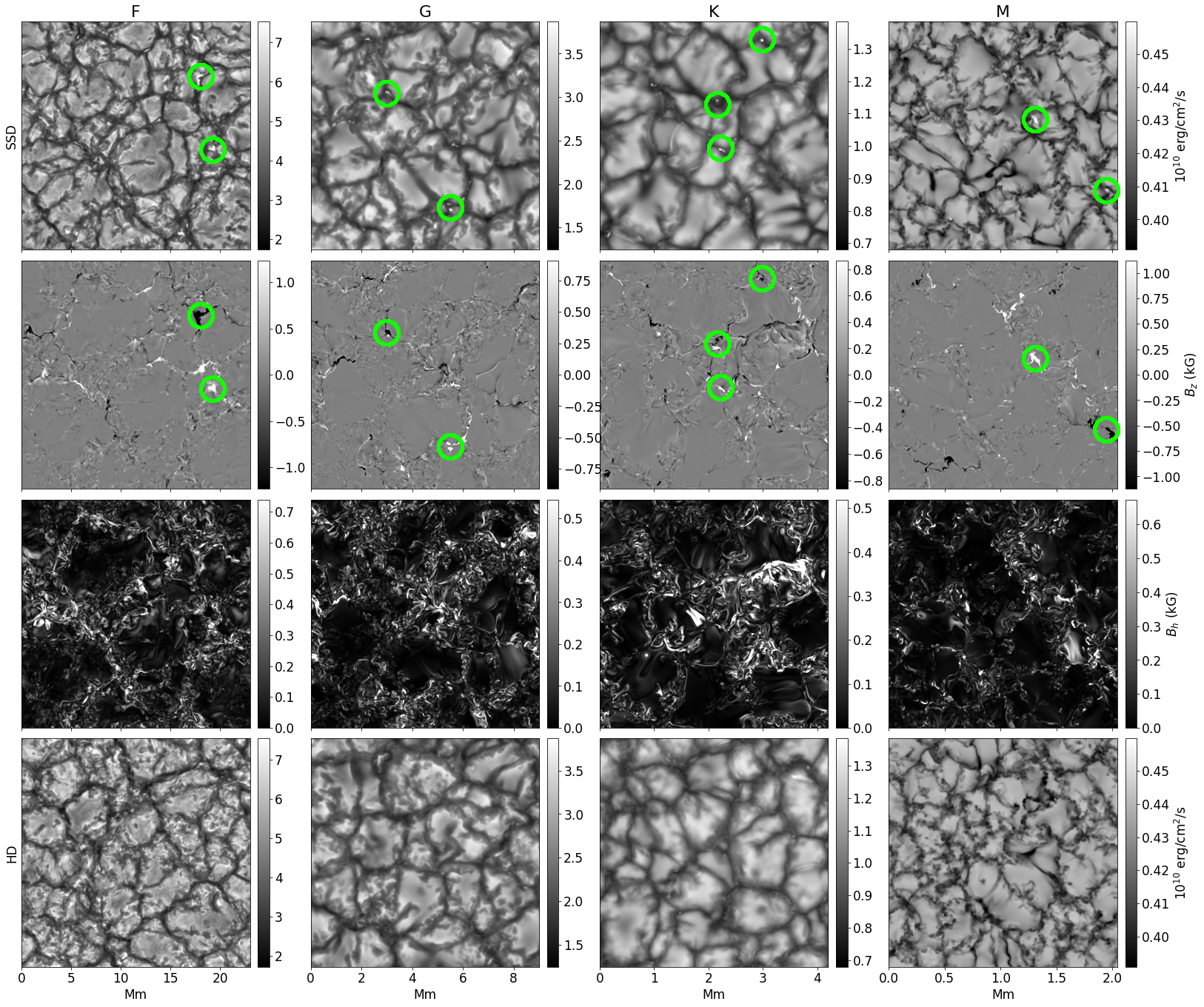}
    \caption{Snapshots of the bolometric intensity $I$ (in $10^{10}$ erg/cm$^2$/s) for the SSD case (\textit{first row}) and the corresponding vertical magnetic field $B_z$ (in kG) (\textit{second row}) and the horizontal magnetic field $B_h$ (in kG) (\textit{third row}) at the iso-$\tau=1$ surface for spectral types (\textit{from left to right}) F, G, K and M, respectively. Snapshots of $I$ for the HD case for comparison (\textit{last row}). The colour bars are identical for the respective SSD and the HD case. Green circles indicate magnetic bright points.}
    \label{fig:snap}
\end{figure*}

\begin{figure}[h]
    \resizebox{\hsize}{!}
    {\includegraphics{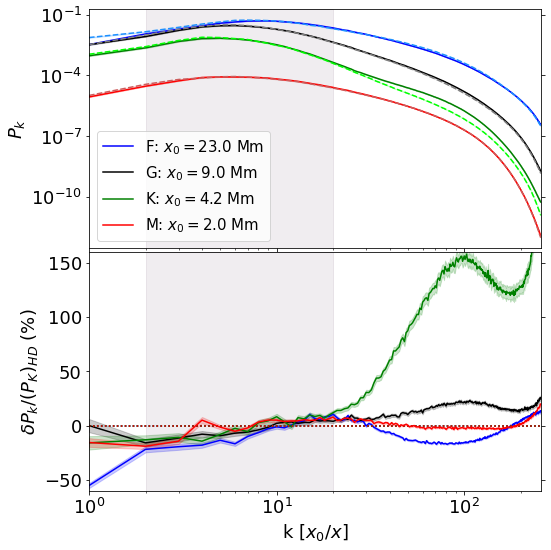}}
    \caption{Spatial power spectra $P_k$ of bolometric intensity $I_{\mathrm{bol}}$ for the SSD (\textit{solid}) and HD (\textit{dashed}) cases plotted against spatial frequency $1/x$ (normalized by the horizontal box size $x_0$ for each star). \textit{Top}: $P_k$ for all cases normalized by the SSD F-star total power ($\sum_k P_k$). \textit{Bottom:} Relative change in power at different scales between the SSD and HD cases. The gray shaded region refer to the approximate scales corresponding to range in granule sizes, for which the centre of gravity is calculated in Table \ref{tab:cog}.}
    \label{fig:i_ps}
\end{figure}

\begin{table*}[h]
    \caption{Average values for various quantities related to $v_z$ and $I_{\mathrm{bol}}$. For each quantity, left column is SSD, right is HD. Here, contrast is defined as $\sigma_I/\avg{I_{\rm bol}}$. All units are cgs.}
    \label{tab:ivzdat}
    \centering
    \begin{tabular}{c|c c | c c |c c |c c |c c}
        \hline\hline
        Simulation & \multicolumn{2}{c|}{upflow frac. [\%]} & \multicolumn{2}{c|}{$\overline{v_{z,\mathrm{rms}}}\, [10^{5}]$ } & \multicolumn{2}{c|}{$\overline{\avg{I_{\mathrm{bol}}}} \, [10^{10}]$} & \multicolumn{2}{c|}{$\overline{\sigma_I} \, [10^{10}]$} & \multicolumn{2}{c}{contrast [\%]}\\
         & SSD & HD & SSD & HD & SSD & HD & SSD & HD & SSD & HD \\
        \hline
        F3V & 57.33&57.14 & 5.167&5.568 & 4.653&4.662 & 1.002&1.026 & 21.54&22.01 \\
        G2V & 56.64&56.83 & 2.518&2.561 & 2.543&2.535 & 0.440&0.444 & 17.30&17.50 \\
        K0V & 55.38&55.69 & 1.316&1.335 & 1.043&1.038 & 0.120&0.121 & 11.51&11.65 \\
        M0V & 60.31&60.05 & 0.632&0.660 & 0.425&0.425 & 0.011&0.011 & 2.63&2.63 \\
        \hline
    \end{tabular}
\end{table*}

The presence of SSD magnetic fields affects the bolometric intensity $I_{\mathrm{bol}}$ in multiple ways. The evacuation of plasma due to concentrated magnetic fields in intergranular lanes leads to formation of bright points. This is generally attributed to the "hot-wall" effect \citep{spruit_brightpts1976}, where the low density plasma in the intergranular lanes is heated up by the surrounding hot, dense upflows, which causes the former to appear bright. This effect is easily seen in the modelled F, G, K and (to a lesser degree) M stars (see Fig. \ref{fig:snap}).

In addition, there are changes in the spatial distribution of $I_{\mathrm{bol}}$ due to the SSD. The top panel of Fig. \ref{fig:i_ps} shows the magnitude of the spatial power spectra $P_k$ of $I_{\mathrm{bol}}$ for all the HD and SSD cases. See appendix \ref{app:ibol_ps} for details on how the power spectrum is calculated. We note that the spatial frequency has been scaled by the box size, which essentially corresponds to a pressure scale height scaling (see Sect. 2.2 of Paper 1 for details). This ensures that all plots have the same range on the $x$-axis. The bottom panel shows the relative change in the power between SSD and HD cases $(P_{k,\mathrm{SSD}}/P_{k,\mathrm{HD}})-1$, with positive values corresponding to an increase in power in the SSD case.

The usual interpretation of $P_k$, as calculated here, is in terms of level of contrast at different spatial scales \citep{nordlund1997_granulation_ps}. We take the peaks of these spectra to be an indication of the granulation scales, as the largest contrast is expected to exist at the scale of granulation (bright granule centres vs. dark intergranular lanes). Due to the variation of granule sizes over a range of scales, we calculate the centre of gravity (CG) over a range of $k$ (marked by the gray shaded region) to estimate an average spatial wave number $k_{\mathrm{CG}}$ corresponding to an average granule size. The peaks of the power spectra lie between $x_0/x \in (4,9)$. With that in mind, we restrict the limits for the CG calculation between $x_0/x \in (2,20)$. For reference, this corresponds to a spatial frequency between 0.22 Mm$^{-1}$ (or 4.5 Mm in length) and 2.2 Mm$^{-1}$ (or 0.45 Mm in length) for the G-star, where the typical granule size is $\sim 1.5-2$ Mm, corresponding to a spatial frequency of $\sim 0.67-0.5$ Mm$^{-1}$. The results are presented in Table \ref{tab:cog}. We interpret the positive change in $k_{\mathrm{CG}}$ for all cases as an indication for a slight decrease in average granule size for SSD cases (as $x \sim 1/k$), relative to the HD cases. The results indicate a positive $\delta k$ (smaller scales) for granulation in the presence of an SSD compared to the HD case. Modifying the limits ($k_1$ from 2 to 3 and $k_2$ from 10 to 20) affects $\delta k$ but not its sign. The value of $\delta k/k_{HD}$ ranges from +1.1\% to +3.8\% for all the stellar types considered (M, K, G and F) for $k_1 = 2$ or 3 and $k_2 = 10$, 15 or 20.

\begin{table}[h]
    \caption{Center-of-gravity (CG) for average granulation scale in Fig. \ref{fig:i_ps}, calculated as $(\int_{k_1}^{k_2}P_k k dk)/(\int_{k_1}^{k_2}P_k dk)$, where $k_1,k_2$ are the left and right bounds of the shaded region.}
    \label{tab:cog}
    \centering
    \begin{tabular}{c|c c c}
        \hline\hline
        Simulation & $k_{\mathrm{CG,SSD}} [x_0/x]$ & $k_{\mathrm{CG,HD}} [x_0/x]$ &  $\delta k/k_{\mathrm{HD}}$ (\%) \\
        \hline
        F3V & 10.5 & 10.08 & +4.09\\
        G2V & 8.65 & 8.41 & +2.77\\
        K0V & 7.57 & 7.30 & +3.68\\
        M0V & 9.36 & 9.15 & +2.24\\
        \hline
    \end{tabular}
\end{table}

\begin{figure}[h]
    \resizebox{\hsize}{!}
    {\includegraphics{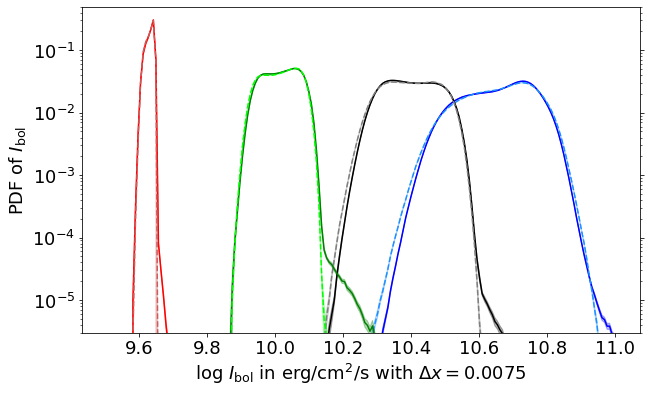}}
    \caption{PDF of the bolometric intensity for the SSD (\textit{solid}) and HD (\textit{dashed}) with vertical axis in semilog.}
    \label{fig:ipdf_semilog}
\end{figure}

In Fig. \ref{fig:ipdf_semilog}, we plot the PDF of the bolometric intensity for all the SSD as well as HD cases. The F, G and K-star show a clearly bimodal distribution, with the bright peak corresponding to mean granular intensities and downflows corresponding to mean intergranular lanes intensities. This is consistent with distributions obtained from a variety of other simulations of stellar photospheres \citep{allcode_beeck2012,stagger1,beeck1,salhab2018}.
The M-star also has a somewhat bimodal distribution, but it is not as prominent as the other cases, since the contrast between granules and intergranular lanes is very low. All SSD cases show an extended bright tail (right side of the PDFs) which corresponds to the formation of magnetic bright points in intergranular lanes. In addition, the G and K-stars show slight excess intensities in the bright flank (above the bright tail, corresponding to bright granules) for the SSD cases, whereas F-star shows a slight decrease. The F and G-stars also show a steeper fall-off at the dark flank (left side of the PDFs), which contributes a reduction in contrast for the SSD cases, as noted in Table \ref{tab:ivzdat}.

The effect of SSD fields on intensity at sub-granular scales is more varied between spectral types. Fig. \ref{fig:i_ps}, bottom panel, shows a prominent increase in power for the K-star at the smallest spatial scales, corresponding to the high-contrast bright points present in the intergranular lanes. This is consistent with the bright tail in Fig. \ref{fig:ipdf_semilog}. A similar interpretation also holds for the G-star. Visually, we also see bright points in the F-star case (see first column in Fig. \ref{fig:snap}). However, these bright points do not lead to an increase in power. This is because magnetic fields also restrict convective velocities, acting as an effective "viscosity", which makes the flow more laminar and leads to granules with a smoother appearance, thus decreasing the amount of small-scale sub-structure and contrast. We caution against interpreting the results in the $k>100$ range, since here the effects of numerical diffusion are non-negligible.

\subsection{Vertical velocity}\label{sec:results:vz}

In Paper 1, we showed that there is a general reduction in vertical velocity $v_z$ for SSD cases near the photosphere. Here we examine in more detail how the distribution of $v_z$ changes relative to the HD cases when accounting for magnetic field generation.

\begin{figure}[h]
    \resizebox{\hsize}{!}
    {\includegraphics{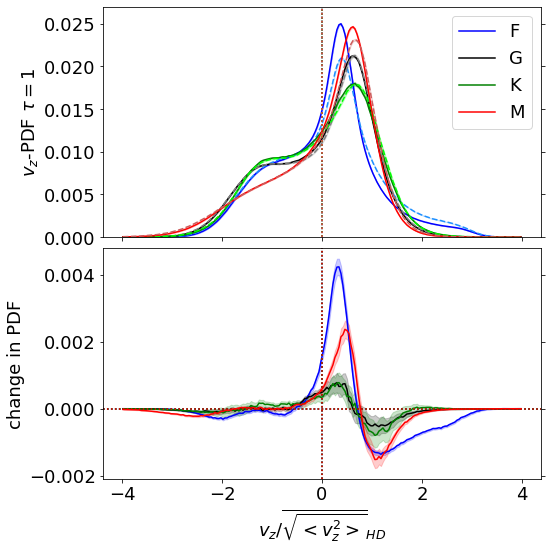}}
    \caption{Vertical velocity $v_z$ at the iso-$\tau=1$ surface normalized by the r.m.s vertical velocity in the HD case $(v_{z_rms})_{\mathrm{HD}}$ (as noted in Table \ref{tab:ivzdat}). \textit{Top}: PDF of $v_z/(v_{z_rms})_{\mathrm{HD}}$ for the SSD (\textit{solid}) as well as HD (\textit{dashed}) cases. \textit{Bottom}: The difference between the PDF for SSD and HD cases $v_{z,\mathrm{SSD}} - v_{z,\mathrm{HD}}$. Positive (negative) values on horizontal axis correspond to upflows (downflows).}
    \label{fig:vz_pdf}
\end{figure}

The top panel of Fig. \ref{fig:vz_pdf} shows the PDF of $v_z$ at the $\tau=1$ iso-surface, normalized by $(v_{z,\mathrm{rms}})_{\mathrm{HD}}$. This normalization allows us to compare the shapes of the PDF between the different stars and to examine the changes between the SSD and HD cases (Fig. \ref{fig:vz_pdf}, bottom panel). First of all, we note that all cases show a similar PDF, with a sharp high peak for upflows and a broad low peak for downflows. There are a couple of exceptions to the general trend: the downflow peak for the M-star is lower than for the others. This might partly be a consequence of a somewhat higher upflow fraction for the M-star ($\sim$60\%) compared to the other stars ($\sim$55-57\%). Another difference is the upflow peak for the F-star, which is offset to relatively smaller velocities compared to the other stellar types. This is consistent with a thicker tail for the high upflow velocities $v_z$, and probably reflects the larger spread in $v_z$ for the F-star. Notwithstanding these small differences, the distribution of velocities is remarkably similar for all the four stars. 

In the presence of SSD magnetic fields, we see that there is a decrease in the mean upflow velocities. This is possibly a consequence of reduced kinetic energy in upflows due to presence of SSD fields, with the fields acting effectively as enhanced viscosity. The mean downflow velocities remain relatively unchanged. This may be related to the fact that magnetic fields in downflows are close to vertical, allowing downflowing plasma to remain relatively unhindered, whereas the magnetic field above granules is largely horizontal, which impedes upflowing plasma \citep{SchVog2008,Rempel2014}.

\subsection{Spatial distribution of energy}\label{sec:results:ener}

\begin{figure}[h]
    \resizebox{\hsize}{!}
    {\includegraphics{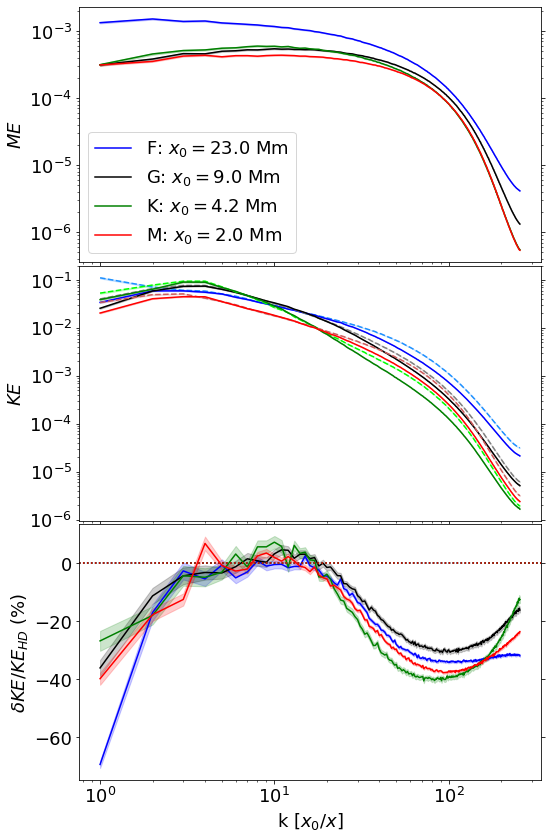}}
    \caption{Spatial power spectrum of magnetic and kinetic energies at the iso-$\tau=1$ surface. \textit{Top}: Power spectrum of magnetic energy for all the SSD cases. \textit{Middle}: Power spectrum of the kinetic energy for all the SSD (\textit{solid, dark}) and HD (\textit{dashed, light}) cases. \textit{Bottom}: Percent change in the kinetic energy power spectrum for SSD cases, relative to HD cases. The top and middle plots are normalized by the total kinetic energy for the F-star (\textit{solid blue}).}
    \label{fig:bke_ps}
\end{figure}

The spatial power spectrum plot for the magnetic energy (ME) in Fig. \ref{fig:bke_ps} (top panel) shows a fairly similar distribution for the G, K and M-star, whereas for the F-star, the spectrum is slightly steeper at the larger scales and has higher power than the spectra for other stars at all wavenumbers. The power spectra for kinetic energy (KE) (middle panel) are also very similar for all the stars at smaller wavenumbers and roughly similar for the larger wavenumbers. In fact, the relative changes in the KE power spectra between the SSD and the HD cases (bottom panel) are remarkably similar for all the models, with a decrease in energy at the largest scales (smallest wavenumbers) and the smallest scales (largest wave numbers). On the other hand, there is no significant change in the power at scales roughly corresponding to granule sizes (see Sect. \ref{sec:results:ibol} for details on granulation scales). Since the dimensions of all the stars are scaled to have a similar number of vertical pressure scale heights (and the horizontal size is scaled accordingly to maintain aspect ratio), the similarity in all the power spectra point to a simple pressure scale height scaling of the relevant dynamics.

\section{Discussion}

\subsection{kG magnetic fields}\label{sec:discuss:bpdf}

\begin{figure*}[h]
    \centering
    \includegraphics[width=17cm]{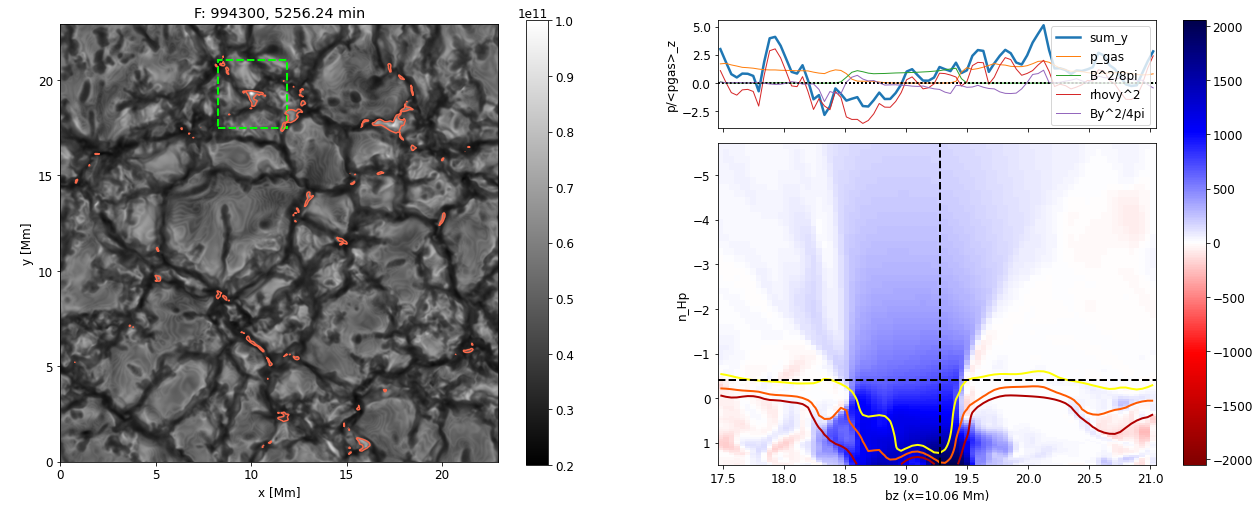}
    \caption{Horizontal force balance in a vertical slice of a magnetic field concentration for the F-star. \textit{Left:} Plot of bolometric intensity with the selected magnetic element highlighted by the green rectangle. Red contours represent areas with $B^2/8\pi>p_{\mathrm{gas}}$. \textit{Right:} Map of $B_z$ in the $x$-plane with horizontal axis in Mm and vertical axis in number of pressure scale heights above (negative) and below (positive) the $\avg{\tau}=1$ height. The dark red, red and yellow contours show the iso-$\log_{10}\tau=$1,0,-1 surfaces, respectively. The line plot on top shows the diagonal terms of the total stress tensor (refer to Eq. \ref{eqn:forces} and Sec. \ref{sec:discuss:bpdf} for details). Animation available.}
    \label{fig:F_pres}
\end{figure*}

\begin{figure*}[h]
    \centering
    \includegraphics[width=17cm]{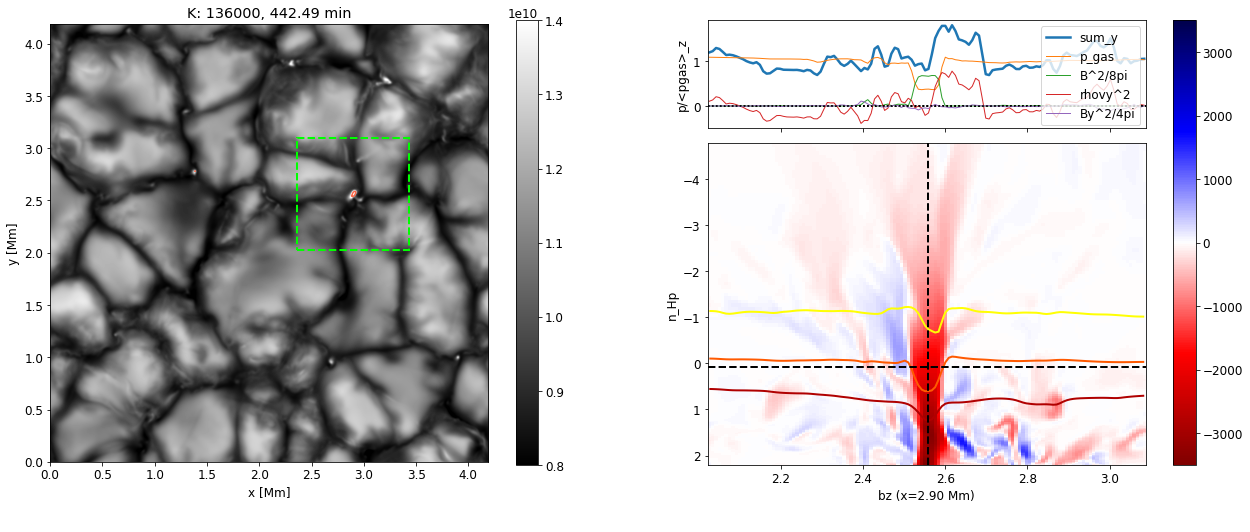}
    \caption{Same as Fig. \ref{fig:F_pres}, but for K-star. Animation available.}
    \label{fig:K_pres}
\end{figure*}

As mentioned in the introduction, the distribution of small-scale magnetic field strengths is well-studied for the solar case. The weak (sub-kG) field distribution is explained in terms of an equipartition between KE and ME: weak turbulent magnetic fields get carried upward and outwards from granule centres and get collected in intergranular lanes up to kinetic energy equipartition in a process called flux expulsion \citep{weiss1996_fe}. However, the field strength corresponding to KE equipartition is substantially sub-kG. To explain the presence of kG fields, the convective collapse mechanism \citep{Par1978,spruit_linear_1979,spruit1979} has usually been invoked.
A qualitative picture of the mechanism is as follows: magnetic field gets accumulated in downflow lanes due to flux expulsion from granule centres. This field can intensify up to equipartition with kinetic energy. At this point, a nascent flux tube forms. While plasma in the tube keeps flowing downwards, the flow of plasma into the lanes is restricted by the Lorentz force, leading to the tube getting evacuated. Finally, to maintain horizontal force balance, the tube gets compressed, causing the magnetic field to amplify, potentially up to pressure equipartition ($p_{\mathrm{mag}} = B^2/8\pi \approx p_{\mathrm{gas}}$).
The efficiency of this mechanism can be thought of in terms of the minimum plasma $\beta = p_{\mathrm{gas}}/p_{\mathrm{mag}}$ (or the strongest fields) for which this instability can work in an idealized flux tube. \citet{rajaguru2002_stellar_cc} showed that this minimum $\beta$ shows an increasing trend with $T_{\mathrm{eff}}$ and decreasing $g$, implying that convective collapse is more efficient for hotter stars.

However, this mechanism can only explain field strength up to pressure equipartition, whereas our simulations reveal presence of super-equipartition fields for the F and G-stars. The reason for this is that convective collapse is an idealized model that assumes flux concentrations exist as stable thin vertical flux tubes and only accounts for the importance of magnetic pressure and gravity. Realistic simulations of flux concentrations have shown that the assumptions are not quite satisfied, especially near the surface where $\beta \sim 1$ \citep{Yel2009}.

\begin{figure}[h]
    \resizebox{\hsize}{!}
    {\includegraphics{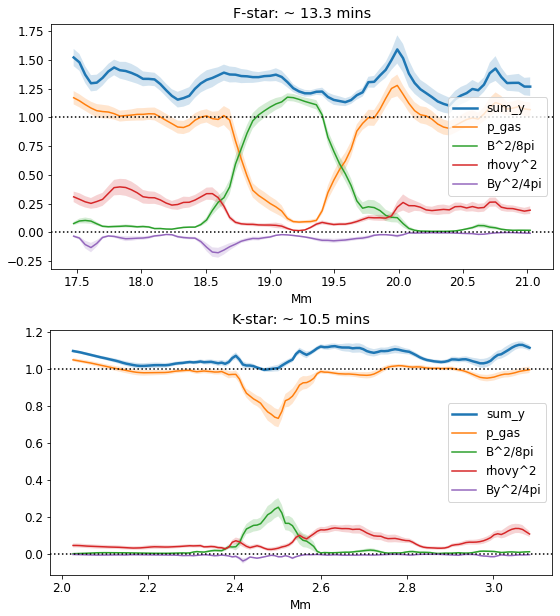}}
    \caption{Average over time of the line plots described in Fig. \ref{fig:F_pres} (\textit{top}) and Fig. \ref{fig:K_pres} (\textit{bottom}), respectively. Error bars are 1$\sigma$ standard error.}
    \label{fig:pbal_horz}
\end{figure}

Some idealized convection studies have indeed shown to result in magnetic field concentrations well above $B_{\mathrm{eqp}}$, for example, in \citet{concflux_bushby2008}, where they cite dynamic pressure as a major factor leading to super equipartition fields. This becomes especially important for hotter stars like our F-star, where velocities in the photosphere are, on average, sonic, and the turbulent pressure is non-negligible compared to gas pressure. To understand which factors are important for the formation and evolution of flux concentrations we consider the momentum balance equation in the following form,
\begin{align}\label{eqn:forces}
    \frac{d (\rho v_i)}{dt} &= -\nabla_j \left( p_{\mathrm{gas}}\delta_{ij} + \rho  v_i v_j +
    \frac{1}{4\pi}\left(\frac{B^2}{2} \delta_{ij}-B_i B_j\right)
    \right)
    + \rho g_i
\end{align}
Here, the indices $i,j$ correspond to the $x,y,z$ directions and $g_i=(0,0,-g_{\mathrm{surf}})$. The terms on the right hand side within the parentheses are the gas pressure and, when averaged over time, the Reynolds and Maxwell stresses, respectively. The Maxwell stresses themselves can be decomposed into an isotropic magnetic pressure term and a general magnetic tension term, respectively.

We consider individual field concentrations for the F-star (Fig. \ref{fig:F_pres}) and the K-star (Fig. \ref{fig:K_pres}). We note that Eq. \ref{eqn:forces} must always be satisfied at any given point of time. For horizontal force balance, however, we only consider the diagonal terms ($i = j$) of the total stress tensor (first bracket on the right hand side) for simplicity. For a cut along the $y$-axis, these terms (gas pressure $p_{\rm gas}$, magnetic pressure $B^2/8\pi$, $\rho v_y^2$ and $B_y^2/4\pi$, plus the sum of all the components) are plotted for the highlighted field concentration in the left panel for both Fig. \ref{fig:F_pres} and \ref{fig:K_pres} at a given instant.

If there was perfect horizontal pressure balance, the thick blue line in Fig. \ref{fig:F_pres} and \ref{fig:K_pres} would be flat. That is clearly not the case, however, even for the K-star. This means we are possibly missing contributions from the time-dependent $d(\rho v_i)/dt$ term, as well as the off-diagonal components ($i \ne j$ in Eq. \ref{eqn:forces}) of the stress tensor. To account for the former term, at least, we average over the lifetime of the magnetic field concentrations for both the F- and K-star. The corresponding averages are plotted in Fig. \ref{fig:pbal_horz}. Luckily, the horizontal balance is reasonably well-maintained (that is, the thick blue line is relatively flat) with just the diagonal components for the F-star (top) as well as the K-star (bottom). The difference between the two cases is the strength of the magnetic pressure relative to the gas pressure: in the F-star, $B^2/8\pi > p_{\rm gas}$ in the centre of the field concentration, whereas for the K-star, $B^2/8\pi \sim p_{\rm gas}/3$. The reason for this difference, in this case, is the extra contribution from the $\rho v_y^2$ term, which is roughly $p_{\rm gas}/4$ outside of the field concentration. We note that the value of $B_{\rm eqp,tot}$ is around 1 kG for the F-star for the $z_{\avg{\tau}=1}$ slice, which explains why the kG fraction in this case is essentially zero.

The degree of evacuation of a flux tube is dependent on the level of superadiabaticity \citep{spruit_linear_1979}, with hotter stars having higher superadiabaticity near the surface \citep{rajaguru2002_stellar_cc,beeck1}. This explains the trend of kG fields in G, K and M-stars relative to pressure-equipartition field strength. However, based on the analysis above, for the F-star, an additional contribution from the diagonal component of the Reynolds stresses (here, $\rho v_y^2$) must be accounted for at the very least. When this is done, the equipartition field strength rises to $\sim 1$ kG and the fraction of field strength above this drops to zero essentially, as noted in section \ref{sec:results:bpdf}. Another thing to note here is the depth of the iso-$\tau=1$ surface in terms of $n_{H_p}$: For the F-star, the dip in the surface in the intergranular lane can be $>1.5 n_{H_p}$, whereas for the K-star, this dip is much less than $1.5 n_{H_p}$. This explains why the PDFs for the iso-$\tau=1$ surface look rather similar for all cases: the horizontal force balance is maintained relatively deeper down for hotter stars (we "see" stronger fields), whereas for cooler stars, the depth difference is not so significant.

\subsection{Granulation and intensity distribution}\label{sec:discuss:ibol}

All models show slight changes in the apparent granulation with the inclusion of SSD fields. In Paper I, we showed that the inclusion of SSD fields results in a reduction in the ratio of horizontal to vertical velocities $v_{h,\mathrm{rms}}/v_{z,\mathrm{rms}}$ as well as in the density scale height $H_{\rho}$. Above the surface, $H_{\rho}$ reduces for the SSD F-star, but stays largely the same for the SSD G, K and M-star, as the decrease in turbulent pressure is roughly compensated by an increase in magnetic pressure. However, $v_{h,\mathrm{rms}}/v_{z,\mathrm{rms}}$ decreases for all cases. Based on simple momentum conservation arguments, the diameters of convection cells at any depth is given by $D \approx 4(v_h/v_z) H_{\rho}$ \citep{nordlund2009}, so one would expect the granule size to decrease accordingly. To check that, we base the average granule size on the peak of the spatial power spectrum $P_k$ of $I_{\mathrm{bol}}$. Previous simulations have shown a tight correlation between granule diameters derived from this relation and $P_k$ \citep{stagger1}. A decrease in granule size would imply a shift to smaller spatial scales (higher spatial frequency) of the peak. This is, in fact, the case for all the stars, as shown in Table \ref{tab:cog}. This is also supported by observational indications of the relation between magnetic field and granule size  for the Sun. Various studies of variation in granule size within an active region \citep{Tit1992,narayan2010} and over the solar activity cycle \citep{ballot2021}, show a general inverse correlation between the granule size and the magnetic field strength.

\begin{figure}[h]
    \resizebox{\hsize}{!}
    {\includegraphics{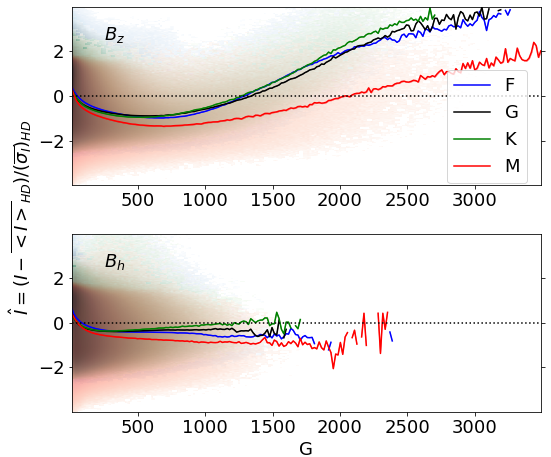}}
    \caption{Mean values of normalized intensity $\hat{I}=(I-\overline{\avg{I}_{\mathrm{HD}}})/\overline{\sigma_{I_{\mathrm{HD}}}}$ for each bin of $B_z$ (\textit{top}) and $B_h$ (\textit{bottom}) at the iso-$\tau=1$ surface, plotted over the respective joint histograms}
    \label{fig:ibz}
\end{figure}

At the smaller scales, the features are more varied between spectral types. As mentioned before, all cases exhibit magnetic bright points in intergranular lanes. We can more clearly see the relation between magnetic field concentrations and intensity by considering the joint histogram of $B_z$ and normalized intensity $\hat{I}=(I-\overline{\avg{I}_{\mathrm{HD}}})/\overline{\sigma_{I_{\mathrm{HD}}}}$ and plotting the mean values of $\hat{I}$ values for each $B_z$ bin (Fig. \ref{fig:ibz}, top panel). Here we see that F, G and K-star have quite similar mean $\hat{I}$ values, with a correlation between bright features and strong $B_z$ values, qualitatively similar to observational findings for the Sun \citep{KahRie2017}. The trend is also similar for the M-star, but the correlation is not as strong. The fact that the plots for F, G and K-star basically overlap is a consequence of the normalization of intensity by $\sigma_I$. In addition, the horizontal fields $B_h$ show no such correlation with intensity (Fig. \ref{fig:ibz}, bottom panel), firmly connecting bright points to vertical field concentrations, confirming the result found for the Sun by, e.g., \citet{RieSol2017}.

The presence of magnetic bright points is expected to enhance power in the $I_{\mathrm{bol}}$ spatial power spectra at smaller scales, but this is clearly the case only for G and K stars. This can be understood in terms of where the $\tau=1$ surface is formed with respect to where the energy transfer shifts from convective to radiative \citep{nordlund1990granulation} for different stellar types. As discussed in Sect. 4.2 of Paper I, for the F and G-star, the $\tau=1$ layer forms below where most of overturning of plasma takes place, leading to naked granules whereas for the K and M-star, it forms above, leading to hidden granules (see also Sect. 3.2 of \citet{beeck2} for a more comprehensive description of what constitutes a naked vs. hidden granule). Especially for the F-star, the turbulent structures within granules are seen very clearly in the HD case. However, with SSD fields, this turbulent appearance smooths out significantly as the magnetic field ends up acting like an effective viscosity hindering the flow. This affects not only the intensity distribution but also the overall radiative flux: For the F-star, there is a slight decrease in the bolometric intensity (see column 4 of Table \ref{tab:ivzdat}), i.e. the decrease in intensity due to the magnetic field's effect on convection dominates over the enhancement due to bright points. This effect is not so strong for the G and K star, since the increase in contrast due to bright points dominates. In addition, there is a slight increase in bolometric intensity. For the M-star, there is practically no change as bright points are relatively infrequent and their contrast is relatively low as well.

\subsection{Energy distribution and vertical velocities}\label{sec:discuss:ener_vels}

Our simulations show remarkably similar spatial distributions of not just KE and ME, but also the change in KE between SSD and HD simulations. Since all boxes are scaled by number of pressure scale heights, this implies a simple pressure scale height proportionality between important scales in the considered stellar atmospheres. The presence of SSD fields results in reduction of KE at sub-granular scales as well as the scale of the whole box. Energy for the magnetic fields is extracted from the KE reservoir at small (sub-granular) scales, which leads to fields with strength near kinetic energy equipartition, and this ME cascades to the largest scales, resulting in a net reduction of average KE. Global simulations of SSD also show a similar reduction in KE at the largest scale \citep{Hotta2022}, pointing to possibly significant global consequences in stellar convection with SSD fields.

An observational signature associated with this reduction in convective velocities is change in the convective blueshift of photospheric lines \citep{Dravins1981}. Convective blueshift is one of the few measurable quantities that encodes information about stellar granulation. A reduction in $v_z$ would imply shifts in line profiles used for estimating convective blueshift. Since these simulations are gray, we refrain from calculating line bisectors etc., as these are likely to be inaccurate. We plan to carry on this analysis in a subsequent paper in this series.

The reduction of KE at smallest scales of around $40\%$ (see Fig. \ref{fig:bke_ps} bottom panel near $k{\sim}10^2$) reflects the near-equipartition division of energy between KE and ME at the scales where field amplification takes place. The SSD mechanism and the corresponding scale-dependent transfer of energy between the kinetic and magnetic energy reservoirs has been extensively studied and was shown to be quite universal \citep{Moll2011}.

\begin{figure}[h]
    \resizebox{\hsize}{!}
    {\includegraphics{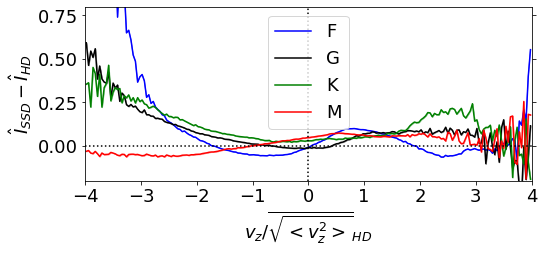}}
    \caption{Difference between the SSD and HD cases for the mean normalized intensity $\hat{I}$ as a function of normalized vertical velocity $v_z/(v_{z,rms})_{\mathrm{HD}}$ at the iso-$\tau=1$ surface. Positive values on the vertical axis mean a higher mean value for the intensity from the SSD case at the corresponding velocity bin. Positive values on the horizontal axis represent upflows.}
    \label{fig:ivz_diff}
\end{figure}

Another way to study the effect of SSD fields on intensity is to consider the joint PDFs of $\hat{I}$ and $v_z/(v_{z,rms})_{\mathrm{HD}}$ and examine the difference between the mean $\hat{I}$ for SSD and HD cases.
In brief, we compute the mean value of intensity in every velocity bin for the SSD and HD case and plot the difference in Fig. \ref{fig:ivz_diff}. We see that, for the F, G and K-star SSD cases, $\hat{I}$ is enhanced in downflows, as would be expected from bright points forming in downflows. Interestingly, there is a slight increase in $\hat{I}$ for upflows as well for all cases (except the F-star), which could mean that granules are in general a little brighter for SSD cases. 

\section{Conclusion}
The presence of SSD fields in our simulations affects the photosphere in a rather similar manner for the cool-star spectral types considered here: fields are amplified due to turbulent plasma motions at sub-granular spatial scales. These fields then get collected in intergranular lanes, where they get concentrated to kG levels while (roughly) maintaining horizontal force balance. For the F-star, the horizontal force balance is satisfied only after inclusion of Reynolds stresses, which leads to magnetic pressure being higher than gas pressure in the strongest flux concentrations. Magnetic bright points are also occasionally visible in the downflow lanes, with a clear correlation of $B_z$ with $\hat{I}$, as well as an increase in the bright tail of intensity distribution. There is also an overall slight decrease in granule size. Because of the SSD fields, the upflow velocities also decrease, again with a similar signature in PDFs of $v_z/(v_{z,rms})_{\mathrm{HD}}$ for all cases. This decrease in upflow velocities signals a possible reduction in expected convective blueshift. In summary, an SSD acting in stellar photospheres is expected to have an effect on spectral line shifts, limb darkening as well as stellar variability at short time scales. We plan to investigate these possibilities in subsequent papers in this series.

\begin{acknowledgements}
    We thank the anonymous referee for their comments and careful consideration, which significantly helped improve the presentation of this paper. TB is grateful for access to the supercomputer Cobra at Max Planck Computing and Data Facility (MPCDF), on which all the simulations were carried out. This project has received funding from the European Research Council (ERC) under the European Union’s Horizon 2020 research and innovation programme (grant agreement No. 695075).
\end{acknowledgements}

\bibliographystyle{bibtex/aa}
\bibliography{bibtex/biblio.bib}

\begin{appendix}
\setlength{\parindent}{0em}

\section{Spatial Power spectrum} \label{app:ibol_ps}
To compute the spatial power spectrum, we use the following procedure:
\begin{enumerate}
    \item For a given 2D quantity $q$, take the 2D FFT (with the zero-frequency mode shifted to the center, e.g., with \texttt{numpy.fft.fftshift} in python using numpy) giving $\Tilde{q}$.
    \item Multiply $\Tilde{q}$ with its complex conjugate $\Tilde{q}^*$ to get $P_{k_x,k_y}=\Tilde{q}\Tilde{q}^*$.
    \item For each radial wave number $k=\sqrt{k_x^2+k_y^2}$, construct a 1-pixel wide mask.
    \item Take the mean of all the data in each mask and multiply by the radius of the mask to get power $P_k$.
\end{enumerate}

For bolometric intensity, we use $q=\sqrt{I}$. For kinetic energy, $q=\sqrt{\rho/2}v$. For magnetic energy, $q=B/\sqrt{8\pi}$.

\section{Additional plots} \label{app:add_plots}

\begin{figure}[h]
    \resizebox{\hsize}{!}
    {\includegraphics{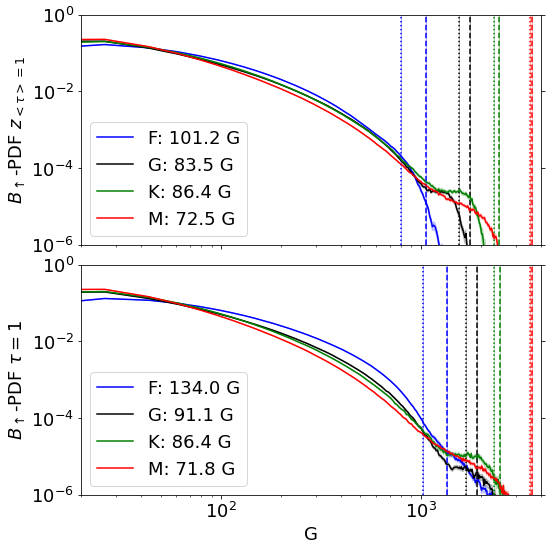}}
    \caption{PDF of the magnitude of magnetic field in upflows, for the geometric surface $z_{\avg{\tau}=1}$ (\textit{top}) and for the $\tau=1$ iso-surface (\textit{bottom}).}
    \label{fig:bpdf_up}
\end{figure}

\begin{figure}[h]
    \resizebox{\hsize}{!}
    {\includegraphics{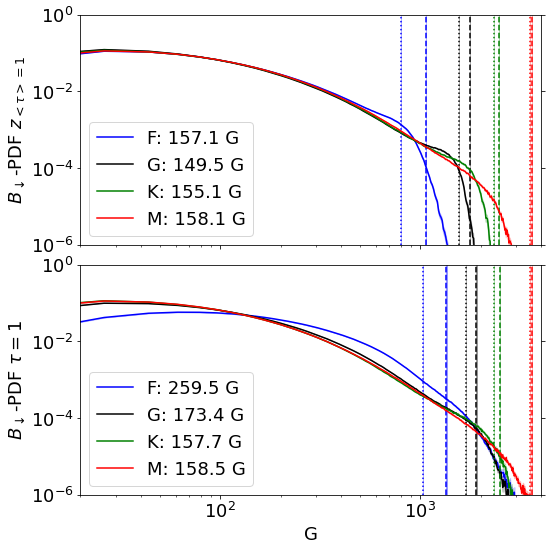}}
    \caption{PDF of the magnitude of magnetic field in downflows, for the geometric surface $z_{\avg{\tau}=1}$ (\textit{top}) and for the $\tau=1$ iso-surface (\textit{bottom}).}
    \label{fig:bpdf_dw}
\end{figure}

\end{appendix}

\end{document}